\documentclass[aps,pra,twocolumn]{revtex4}
\usepackage{graphicx}
\usepackage{dcolumn}
\usepackage{bm}
\usepackage{amsmath}
\usepackage{amssymb}

\allowdisplaybreaks

\def\tr{\mbox{tr}}

\newtheorem{lemma}{Lemma}
\newtheorem{corollary}{Corollary}

\begin{document}

\title{Complete analysis for three-qubit mixed-state discrimination}

\author{Donghoon Ha}
\author{Younghun Kwon}
\email{yyhkwon@hanyang.ac.kr}

\affiliation{Department of Physics, Hanyang University, Ansan,
Kyunggi-Do, 425-791, South Korea}

\date{\today}

\begin{abstract}
In this letter, by treating minimum-error state discrimination as a
complementarity problem, we obtain the geometric optimality
conditions. These can be used as the necessary and sufficient
conditions to determine whether every optimal measurement operator
can be nonzero. Using these conditions and an inductive approach, we
demonstrate a geometric method and the intrinsic polytope for $N$-qubit-mixed-state discrimination. When the intrinsic polytope
becomes a point, a line segment, or a triangle, the guessing
probability, the necessary and sufficient condition for the exact
solution, and the optimal measurement are analytically obtained. We
apply this result to the problem of discrimination to arbitrary three-qubit mixed states with given {\it a priori} probabilities and obtain the
complete analytic solution to the guessing probability and optimal
measurement.
\end{abstract}

\maketitle

 The goal of quantum-state
discrimination is to distinguish between states of a given set as
well as possible. In other words, it can be regarded as a problem to
find the optimal measurement for discriminating among the given
quantum states. In fact, every state in classical physics can be
orthogonal to each other and therefore distinguished
perfectly\cite{ref:chef1}. However, in  quantum physics, a state
cannot be perfectly discriminated because of the existence of
nonorthogonal states \cite{ref:hels1, ref:hole1, ref:barn1}.
Quantum-state discrimination\cite{ref:berg1} is classified into
minimum error discrimination, originally introduced by Helstrom
\cite{ref:hels1}, unambiguous discrimination\cite{ref:ivano2,
ref:diek1, ref:pere2}, and  maximum confidence
discrimination\cite{ref:crok1}. The purpose of minimum error
strategy is to find the optimal measurement and the minimum error
 probability (or guessing probability) for arbitrary $N$-qudit mixed quantum states with arbitrary {\it a priori} probabilities.
 In the $N=2$ case, regardless of the dimension, the Helstrom bound
\cite{ref:hels1} gives an analytic solution to the problem. In the $N=3$
case the analytic solution for pure qubit states is provided by
\cite{ref:hunt1, ref:sams1}. In \cite{ref:jafa1} the analytic solution for
mixed qubit states is considered without the necessary and
sufficient conditions for the solution. In other words, the full
understanding for discrimination of three-qubit mixed quantum states is
not provided yet.

The optimal measurement for linearly independent quantum states is
the von Neumann measurement \cite{ref:eldar1}. But if the given quantum states are linearly dependent, the von Neumann measurement may not be optimal. Therefore, the positive-operator-valued-measure (POVM) should be used for arbitrary quantum states. From the point where POVM can be used as a measurement and the probability to guess the quantum states correctly becomes convex, the minimum error discrimination problem may be solved  by convex optimization  \cite{ref:boyd1}. There have also been some efforts to solve it using a dual problem \cite{ref:deco1}
  or complementarity problem \cite{ref:bae1}.
By applying qubit-state geometry to the optimality conditions for
measurement operators and complementary states, Bae \cite{ref:bae2}
obtained a geometric method to find the guessing probability and the
optimal measurement for some special cases. However, they did not
comprehend the case where optimal measurement cannot be POVM, whose
every element is nonzero. In this article, by showing that the case where optimal measurement cannot be POVM, whose
every element is nonzero, can be understood through the parameters
satisfying the geometric optimality conditions\cite{ref:bae1}, we clarify the
meaning of the geometric condition. Through the 
conditions and the inductive approach, we propose a method to
discriminate arbitrary $N$-qubit mixed quantum states with arbitrary
{\it a priori} probabilities. In this method, we define the intrinsic
polytope for discrimination problems. When the polytope becomes a
point, line segment, or triangle, we find the guessing probability,
the necessary and sufficient condition for the exact solution, and
the optimal measurement analytically. By the number of extreme
points for the intrinsic polytope and the geometric optimality
conditions, we can provide a complete analysis for the
discrimination of the three-qubit mixed state. We also obtain its
guessing probability and optimal measurement.

Let $q_{i}$ and $\rho_{i}$($i=1, \cdots ,N)$ be the {\it a priori} probability and $d \times d$ the density matrix, where $d$ and $N$ denote the dimension and number of states to be discriminated. Hereafter, $q_{i}$ is ordered by $q_{i} \geq q_{i+1}$.  When $\{M_{i}\}_{i=1}^{N}$ is used for measurement to $\{q_{i},\rho_{i}\}_{i=1}^{N}$,
 the probability to guess the quantum states correctly becomes $P_{\rm{corr}}=\sum_{i=1}^{N}q_{i}\tr\rho_{i}M_{i}$. The goal of the minimum error state discrimination is to obtain the maximum of $P_{\rm{corr}}$, called {\it the guessing
probability} $P_{\rm{guess}}$, using POVM. Therefore, the minimum error state discrimination can be described as
\begin{eqnarray}\label{eq:mecov}
&\max& \sum_{i=1}^{N} q_{i} \tr\rho_{i}M_{i}, \nonumber\\
&\rm{subject}\ \rm{to}&  M_{i}\geq 0\quad \forall i\in \{1, \cdots, N\}, \nonumber\\
& &\sum_{i=1}^{N} M_{i} = I_{d} .  \label{eq:convex}
\end{eqnarray}
By semidefinite programming\cite{ref:boyd1}, the dual problem of Eq.(\ref{eq:convex}) is obtained as follows:
\begin{eqnarray}
& \min & \tr K, \nonumber\\
&\rm{subject}\ \rm{to}& K-q_{i}\rho_{i} \geq 0  \quad \forall i\in \{1, \cdots, N\},\label{eq:dual}
\end{eqnarray}
where $K$ is the $d\times d$ Hermitian matrix. In fact, using a
non-negative number $r_{i}$ and the density matrix $\tilde{\rho}_{i}$,
the constraints of the dual problem can be written as
\begin{equation}
K=q_{i}\rho_{i} + r_{i}\tilde{\rho}_{i}\quad  \forall i\in \{1, \cdots, N\}.
\end{equation}
Since the above operator $K$ is equal for all $i$, the following relation
holds
\begin{equation}\label{eq:dualcont}
q_{i}\rho_{i}-q_{j}\rho_{j}=r_{j}\tilde{\rho}_{j}-r_{i}\tilde{\rho}_{i}\quad  \forall i,j\in
\{1, \cdots, N\} .
\end{equation}
In the optimization problem, the dual problem in general has {\it weak duality} and may not be identical to the original one. However,
if the optimization problem is convex and satisfies Slater's
condition, the dual one has  {\it strong duality} and is
equivalent to the primal one. This condition is to check whether
every POVM element is nonzero. Therefore, our problem is equivalent to the
following:
\begin{eqnarray}
& \min & q_1 +r_1, \nonumber\\
&\rm{subject}\ \rm{to} &r_{i}\tilde{\rho}_{i}-r_{j}\tilde{\rho}_{j}=q_{j}\rho_{j}-q_{i}\rho_{i}\quad \forall i,j.
\end{eqnarray}
The objective function can be $q_{i}+r_{i}$$(i=2,\cdots,N)$ instead of $q_{1}+r_{1}$.
By considering the Karush-Kuhn-Tucker (KKT) conditions, let us
investigate the necessary conditions of
$\{M_{i},r_{i},\tilde{\rho}_{i}\}_{i=1}^{N}$, which satisfy
 $P_{\rm corr}=q_{1}+r_{1}$. These conditions contain the constraints of the primal and dual problems as well as the complementary slackness one. The final condition can be found by
 connecting the measurement operators $\{M_{i}\}_{i=1}^{N}$ and $\{r_{i},\tilde{\rho}_{i}\}_{i=1}^{N}$, which are complementary to the constraints of the primal and dual problems:
\begin{equation}\label{eq:compc}
r_{i} \tr[\tilde{\rho}_{i} M_{i}]=0
\quad \forall i\in \{1,\cdots,N\}.
\end{equation}
The KKT conditions, summarized in the following, can be
 derived from the POVM constraints and the no-signaling ones \cite{ref:bae3}:
\begin{eqnarray}\label{eq:kktc}
&\mbox{(i)}&M_{i} \geq 0 \ \mbox{and}\ \sum_{i=1}^{N} M_{i}=I_{d}\quad \forall i, \nonumber\\
&\mbox{(ii)}&r_{i}\tilde{\rho}_{i}-r_{j}\tilde{\rho}_{j}=q_{j}\rho_{j}-q_{i}\rho_{i}\quad \forall i,j, \nonumber\\
&\mbox{(iii)}&r_{i}\tr[\tilde{\rho}_{i} M_{i}]=0\quad \forall i.
\end{eqnarray}
We now obtain the guessing probability and the optimal measurement,
by only these conditions. The complementarity problem is the one where a solution is found for the optimization problem by using the optimality conditions which should satisfy the parameters of the primal and dual problem. In this article  $*$ is used to denote the optimality of the parameters.

Henceforth, by confining only the case of the
two-level system ($d=2$) let us obtain the geometric condition for
Eq. (\ref{eq:kktc}). From the Bloch representation
$\rho_{i}=\frac{1}{2}(I_{2}+\vec{v}_{i}\cdot\vec{\sigma})$ and
$\tilde{\rho}_{i}=\frac{1}{2}(I_{2}+\vec{w}_{i}\cdot\vec{\sigma})$
we can derive the following relations:
\begin{eqnarray}
q_{i}-q_{j} &=& r_{j}-r_{i},\label{eq:dc1}\\
q_{i}\vec{v}_{i}-q_{j}\vec{v}_{j} &=&
r_{j}\vec{w}_{j}-r_{i}\vec{w}_{i}\quad \forall i,j\in\{1,
\cdots, N\},
\end{eqnarray}
where $\vec{v}_{i}$ and $\vec{w}_{i}$ are the Bloch vectors and
$\vec{\sigma}$ represents the Pauli matrices. Since we assume $q_{i} \geq
q_{i+1}$, we can find $r_{i}^* \leq r_{i+1}^*$ from Eq. (\ref{eq:dc1}). Therefore if
$r_{1}^* \neq 0$,  we have $r_{i}^* > 0 $ $(i=1,\ldots,N)$.
 Here, let us take an inductive approach to $N$-qubit-state discrimination,
which means that by assuming that the way to discriminate ($N-1$)
states may be known, we investigate a method to discriminate among the $N$-qubit states. Therefore it is sufficient to consider only those cases where every optimal POVM element is nonzero. (For generality, we will later consider cases  where some of the optimal POVM elements may be zero.) First, we consider cases where every optimal POVM element is nonzero and the guessing probability is greater than
$q_{1}$. In this case, since $r_{1}^*$ is nonzero, the condition
(iii) becomes $\tr(\tilde{\rho}_{i}M_{i})=0$, which implies that the
rank of $\tilde{\rho}_{i}$ and $M_{i}$ should be one. This means that
for each $i$, we find $||\vec{w}_{i}||_{2}=1$ and
\begin{equation}\label{eq:optm}
 M_{i} =p_{i}(I_{2}-\vec{w}_{i}\cdot \vec{\sigma} ),\ p_i >0.
\end{equation}
Since $\{M_{i}^*\}_{i=1}^{N}$ is POVM,
$\{p_{i},\vec{w}_{i}\}_{i=1}^{N}$ should satisfy
\begin{equation}\label{eq:pwc}
\sum_{i=1}^{N} p_{i}\vec{w}_{i}=0,\ \sum_{i=1}^{N} p_{i}=1.
\end{equation}
 Therefore, $\{r_{i},\vec{w}_{i}\}_{i=1}^{N}$ is necessary to satisfy the following conditions (which we will call the geometric KKT conditions):
\begin{eqnarray}
&\mbox{(i)}&  \ r_{i} \vec{w}_{i} - r_{j} \vec{w}_{j} = q_{j}\vec{v}_{j}-q_{i}\vec{v}_{i} \quad\forall i,j, \nonumber\\
&\mbox{(ii)}&  \exists \ \left\{p_{i}\right\}_{i=1}^{N}\ \mbox{s.t.}\ p_{i}> 0\ \forall i,\ \sum_{i=1}^{N} p_{i}\vec{w}_{i}=0,\ \sum_{i=1}^{N} p_{i}=1 ,\nonumber\\
&\mbox{(iii)}&  \parallel \vec{w}_{i}\parallel_{2}=1 \quad\forall i, \nonumber\\
&\mbox{(iv)}& r_{i} - r_{j} = q_{j} - q_{i} \quad\forall i,j .\label{eq:nonzerogeo}
\end{eqnarray}
 Next, we will show that even when every optimal POVM element is nonzero and the guessing probability
becomes $q_{1}$, $\{ r_{i},\vec{w}_{i} \}_{i=1}^{N}$ is necessary to satisfy the above condition Eq. (\ref{eq:nonzerogeo}). We will prove this by considering both cases $q_{1}=q_{2}$ and $q_{1} > q_{2}$. The
case of $q_{1}=q_{2}$ implies $\rho_{1}=\rho_{2}$ by the KKT
condition (ii), which turns out to be the case of discriminating
among the same quantum states. However, we may exclude this case since we
are interested in discriminating entirely different quantum states.
In the case of $q_{1} > q_{2}$, we can see that since $r_{2}^{*} >
0$, $\{ q_{1}-q_{i},(q_{1}\vec{v}_{1} - q_{i}\vec{v}_{i}
)/(q_{1}-q_{i}) \}_{i=2}^{N}$ satisfies the geometric KKT conditions
(i), (iii), and (iv). If $r_{1}=0$, the geometric conditions
(i) and (iv) do not put any restriction on $\vec{w}_{1}$. In addition
$\vec{w}_{1}$ satisfies $\parallel
\vec{w}_{1}\parallel_{2}=1$ and the geometric KKT condition (ii).
From these facts we can see that $\{
r_{i},\vec{w}_{i} \}_{i=1}^{N}$ should satisfy every geometric KKT
condition.

Until now we showed that if every optimal POVM element is nonzero,
we can find $\{r_{i},\vec{w}_{i}\}_{i=1}^{N}$ by satisfying the
geometric KKT conditions. Now we will prove the reverse. That is, we
will prove that if $\{r_{i},\vec{w}_{i}\}_{i=1}^{N}$
satisfies the geometric KKT conditions, every optimal POVM element
can be nonzero. For this let us assume that $\{r_{i},\vec{w}_{i}\}_{i=1}^{N}$ satisfies the geometric KKT
conditions. When $\vec{R}\equiv q_i \vec{v}_i +r_i \vec{w}_i
(i=1,\ldots,N)$, the following relation holds:
\begin{eqnarray}
\sum_{i=1}^{N} q_i p_i (1-\vec{v}_i \cdot \vec{w}_i)
&=& \sum_{i=1}^{N} (q_1 +r_1 -r_i )p_i - \sum_{i=1}^N q_{i} p_{i} \vec{v}_i \cdot \vec{w}_i  \nonumber\\
&=&(q_1 +r_1 )- \sum_{i=1}^N r_i p_i  \| \vec{w}_i \|_2 ^2   \nonumber\\
&&- \sum_{i=1}^N q_{i} p_{i} \vec{v}_i \cdot \vec{w}_i  \nonumber\\
&=&(q_1 +r_1) -\sum_{i=1}^N p_i \vec{w}_i \cdot \vec{R}  \nonumber\\
&=& q_1 +r_1 .
\end{eqnarray}
Then by $\{M_{i}\}_{i=1}^{N}$ given in Eq.(\ref{eq:optm}), we can see that
$P_{\rm corr}$ of the primal problem is equal to
 $q_{1}+r_{1}$:
\begin{equation}
P_{{\rm corr}}=\sum_{i=1}^{N} q_{i} \tr \rho_{i} M_{i} = \sum_{i=1}^{N} q_{i} p_{i}  (1-\vec{v}_{i}\cdot \vec{w}_{i}) =q_1 +r_1.
\end{equation}
Therefore $\{M_{i},r_{i},\vec{w}_{i}\}_{i=1}^{N}$ become the optimal parameters of our primal and
dual problems. Since every $p_{i}$ is positive we can see that all
the POVM elements are nonzero. From this, the following lemma \ref{lem:gk} can be obtained.
\begin{lemma}[geometric KKT conditions]\label{lem:gk}
The fact that every optimal POVM element can be nonzero is
equivalent to the fact that $\{r_{i},\vec{w}_{i}\}_{i=1}^{N}$
satisfying the geometric KKT conditions exists.
\end{lemma}
Let us denote $P\{\vec{x}_{i}\}_{i=1}^{N}$ as the polytope formed by $\{\vec{x}_{i}\}_{i=1}^{N}$. When the number of extreme points of $P \{q_i , \rho_i \}_{i=1}^N (\equiv P\{q_{i}\vec{v}_{i}\}_{i=1}^{N} )$ is the same as the number of
quantum states to be discriminated, the geometric meaning of Eq.(\ref{eq:nonzerogeo}) can be easily expressed.  Then the geometric condition (i) indicates
that $P\{q_{i},\rho_{i}\}_{i=1}^{N}$ is congruent to $P\{
 r_{i}\vec{w}_{i}\}_{i=1}^{N}$. The
 geometric condition (ii) implies that the origin of the Bloch
 sphere lies in the relative interior of $P\{ r_{i}\vec{w}_{i}\}_{i=1}^{N}$. The
 geometric condition (iii) ensures that the distances from the
 origin to the extreme points of
$P\{ r_{i}\vec{w}_{i}\}_{i=1}^{N}$ become $\{
r_{i}\}_{i=1}^{N}$. The final condition
 (iv) shows that the difference between the distances should be the same as that between the {\it a priori} probabilities.  Since $\{r_{i},\vec{w}_{i}\}_{i=1}^{N}$ satisfying the geometric
 KKT conditions (i)--(iii) certainly exists, the crucial element for obtaining the guessing probability is condition (iv).

Let us explain how to discover the guessing probability when
$\{r_{i},\vec{w}_{i}\}_{i=1}^{N}$ cannot satisfy the geometric KKT
conditions. In this case at least one of the optimal POVM elements is
zero. Therefore, if we denote $P_{\rm
guess}^{(N)}(\{q_{i},\rho_{i}\}_{i=1}^{N})$ the guessing probability
function for $N$-qubit states, we may write it as
\begin{equation}\label{eq:zdogp}
P_{\rm guess} = \max_{S} \left(\mbox{$\sum_{j \in S} q_j $}\right)  P_{\rm guess}^{(|S|)} \left( \left\{\mbox{$q_i / \sum_{j \in S} q_j , \rho_i $} \right\}_{i \in S} \right),
\end{equation}
where $S$ is the proper subset of $\{ 1,\cdots, N \}$. For now,
using $S$ and lemma \ref{lem:gk}, we will obtain the guessing probability and
the optimal measurement when $P\{q_{i},\rho_{i}\}_{i=1}^{N}$ becomes
a special case. First, let us consider when
$P\{q_{i},\rho_{i}\}_{i=1}^{N}$ becomes a point. For this purpose,
suppose that $\{r_{i},\vec{w}_{i}\}_{i=1}^{N}$ satisfies the
geometric KKT conditions. Then the conditions (i) and (iii) imply
the equality of $\vec{w}_{i}(i=1,\cdots,N)$. Applying this result to condition (ii), we find that $\sum_{i=1}^{N}p_{i}=0$ and
$\sum_{i=1}^{N}p_{i}=1$, which contradicts each other. Therefore we can see
that when $P\{q_{i},\rho_{i}\}_{i=1}^{N}$ forms a point, every
optimal POVM element cannot be nonzero. Since for any proper subset
$S$ of $\{1,\cdots,N \}$, $P\{q_{i}/\sum_{j \in S} q_{j},\rho_{i}\}_{i\in S}$ becomes a point, the nonzero element of
the optimal POVM is only one. Therefore, we find corollary \ref{coro:ep1}.
\begin{corollary}\label{coro:ep1}
If the number of the extreme points to $P\{q_{i} , \rho_{i}\}_{i=1}^{N}$ is one,  every optimal POVM element except
$M_{1}$  is zero, and the guessing probability is $q_{1}$.
\end{corollary}
 The second case is when $P\{q_{i},\rho_{i}\}_{i=1}^{N}$ forms a line segment. Let us denote the two indices corresponding to
  the extreme points as $\alpha$ and $\beta (>\alpha)$. Then the geometric KKT condition (i) indicates that $P\{r_{i}
\vec{w}_{i}\}_{i=\alpha,\beta}$ should be a line segment with the same
length to $P\{q_{i},\rho_{i}\}_{i=\alpha ,\beta}$. Condition (ii) requires that $P\{r_{i}
\vec{w}_{i}\}_{i=\alpha,\beta}$ contain the origin $O$. This implies
that the length of the line segment becomes
\begin{eqnarray}
& & r_{\alpha} \| \vec{w}_{\alpha}\|_{2}+
r_{\beta} \| \vec{w}_{\beta}\|_{2} =
 \| q_{\alpha}\vec{v}_{\alpha}-q_{\beta}\vec{v}_{\beta} \|_{2} \nonumber\\
&=& r_{\alpha} + r_{\beta}.\label{eq:tsk3}
\end{eqnarray}
The equality in the second line comes from the condition (iii).
Also, by applying condition (iv) to Eq.(\ref{eq:tsk3}) we have
\begin{eqnarray}
r_{\alpha}&=&\frac{1}{2}(\|q_{\alpha}\vec{v}_{\alpha}-q_{\beta}\vec{v}_{\beta}
\|_{2}-(q_{\alpha}-q_{\beta})) \nonumber\\
r_{\beta}&=&\frac{1}{2}(\|
q_{\alpha}\vec{v}_{\alpha}-q_{\beta}\vec{v}_{\beta}
\|_{2}+(q_{\alpha}-q_{\beta})) \nonumber\\
\vec{w}_\alpha &=& \frac{q_\alpha \vec{v}_\alpha - q_\beta \vec{v}_\beta}{\| q_\alpha \vec{v}_\alpha - q_\beta \vec{v}_\beta \|_2} , \vec{w}_\beta = \frac{  q_\beta \vec{v}_\beta - q_\alpha \vec{v}_\alpha}{\| q_\alpha \vec{v}_\alpha - q_\beta \vec{v}_\beta \|_2}
\end{eqnarray}
Since $r_{\alpha},\ r_{\beta}$ should be non-negative, we find $\|
q_{\alpha}\vec{v}_{\alpha}-q_{\beta}\vec{v}_{\beta} \|_{2} \geq
q_{\alpha}-q_{\beta}$. It supplies the necessary and sufficient
condition for $\{r_i , \vec{w}_i \}_{i=\alpha, \beta}$ to satisfy
the geometric KKT conditions. If $\{q_i ,
\rho_i\}_{i=\alpha,\beta}$ satisfies the condition, the guessing
probability becomes
\begin{eqnarray}
P_{\rm guess}&=&\frac{1}{2}((q_\alpha +r_{\alpha})+ (q_\beta + r_{\beta})) \nonumber\\
&=& \frac{1}{2}(q_\alpha +q_\beta+\| q_{\alpha}\vec{v}_{\alpha}-q_{\beta}\vec{v}_{\beta} \|_{2})\nonumber\\
&=& \frac{q_{\alpha} +q_{\beta}}{2}\left[1+ \left\| \frac{q_\alpha \vec{v}_{\alpha}}{q_{\alpha} +q_{\beta}}-\frac{q_\beta \vec{v}_{\beta}}{q_{\alpha} +q_{\beta}} \right\|_{2} \right]. \label{eq:eptgp}
\end{eqnarray}
 From this result, our problem can be thought as one of discriminating $\{q_{i}/(q_\alpha +q_\beta) ,
\rho_i\}_{i=\alpha,\beta}$, with the probability
$(q_{\alpha}+q_{\beta})$. However, if the condition does not hold, we
have to find the index set $S$ which provides the guessing
probability given by Eq. (\ref{eq:zdogp}). However, by this assumption,
since for any $S$ $P\{q_{i}/\sum_{j \in S} q_{j},\rho_{i}\}_{i \in
S}$ forms a point or a line segment, the problem becomes how to discriminate two
quantum states. From the Helstrom bound, we can obtain corollary \ref{coro:ep2}.
 \begin{corollary}\label{coro:ep2}
If the number of the extreme points to
$P\{q_{i} , \rho_{i}\}_{i=1}^{N}$ is two, the guessing probability
becomes
\begin{equation}
P_{\rm{guess}}=\max_{i\neq j}  \frac{1}{2}\left( q_i +q_j +\left\| q_i \rho_i  -q_j \rho_j  \right\|_1 \right).\label{eq:ep2gp}
\end{equation}
 When $a$ and $b( > a)$ are the indices giving the optimal value,
if $\left\| q_a \vec{v}_a  -q_b \vec{v}_b \right\|_2
 < q_a - q_b$, every optimal POVM element except $M_1$ is zero.
However, if $\left\| q_a \vec{v}_a  -q_b \vec{v}_b \right\|_2
 \geq q_a - q_b$, the optimal POVM elements are given as
\begin{eqnarray}
M_a &=& \frac{1}{2} \left[I_2 + \left(\frac{q_a \vec{v}_a - q_b \vec{v}_b}{\| q_a \vec{v}_a - q_b \vec{v}_b\|_2} \right)\cdot \vec{\sigma} \right], \nonumber\\
M_b &=&  \frac{1}{2} \left[I_2 + \left(\frac{q_b \vec{v}_b - q_a \vec{v}_a}{\| q_a \vec{v}_a - q_b \vec{v}_b\|_2} \right)\cdot \vec{\sigma} \right], \nonumber\\
M_i &=&0 \quad \forall i \neq a,b.
\end{eqnarray}
\end{corollary}
 Now let us consider the case when $N=3$, and the intrinsic polytope
forms a triangle. We define two sides of the triangle as
\begin{eqnarray}\label{eq:tlen}
l_{1}&\equiv& \parallel q_{2}\vec{v}_{2}-q_{1}\vec{v}_{1} \parallel_{2}, \nonumber\\
l_{2}&\equiv& \parallel q_{3}\vec{v}_{3}-q_{1}\vec{v}_{1} \parallel_{2},
\end{eqnarray}
and the difference between the {\it a priori} probabilities as
\begin{equation}
e_{1}\equiv q_{1}-q_{2},\ e_{2}\equiv q_{1}-q_{3}.
\end{equation}
 Now suppose that $\{r_i , \vec{w}_i \}_{i=1}^{N}$ satisfies the geometric KKT conditions.
 In this case the number of extreme points is equal to that of
the quantum states to be discriminated. Then
$P\{r_{i}\vec{w}_{i}\}_{i=1}^{3}$ is congruent to
$P\{q_{i},\rho_{i}\}_{i=1}^{3}$, and the origin $O$ exists inside the
relative interior. When $T_{i}$ represents the vertex
$r_{i}\vec{w}_{i}$ of the triangle $P\{r_{i}\vec{w}_{i}\}_{i=1}^{3}$
and $r_{i}$($i=1,2,3$) is the distance from $O$ to the vertex
$T_{i}$, we have the following relations:
\begin{equation}
r_{2}-r_{1}=e_{1},\ r_{3}-r_{1}=e_{2}.
\end{equation}
The necessary and sufficient condition that $\{r_{i},\vec{w}_{i}\}_{i=1}^{3}$,
satisfying that the geometric KKT conditions can exist, can be
obtained by the property of hyperbola, as follows:
\begin{eqnarray}\label{eq:tqnsc}
&{\rm(i)}& l_{1} > e_{1},\ l_{2} > e_{2},\nonumber \\
&{\rm(ii)}&\frac{l_{1}\cos\theta_{1} +e_{1}}{l_{1}+e_{1}} < \frac{l_{1} - e_{1}}{l_{2}-e_{2}},
\  \frac{l_{2} \cos \theta_{1}+e_{2}}{l_{2}+e_{2}} < \frac{l_{2}-e_{2}}{l_{1}-e_{1}},\nonumber \\
&{\rm(iii)}&\frac{l_{1}^{2}-e_{1}^{2}}{2(l_{1}\cos\chi +e_{1})} <  \frac{l_{1}\sin\theta_{2}}{\sin(\chi+\theta_{2})},
\end{eqnarray}
where $\theta_{i}$ denotes the inside angle of
vertex $T_i$, and the angle $\chi$ which is $\angle OT_1 T_2$, is given as
\begin{widetext}
\begin{eqnarray}
\chi&=& \chi_{2} -\chi_{1}, \nonumber\\
\chi_{1}&=&
\cos^{-1}\left(\frac{l_{1}(l_{2}^{2}-e_{2}^{2})-l_{2}(l_{1}^{2}-e_{1}^{2})\cos\theta_{1}
}{\sqrt{l_{1}^{2}(l_{2}^{2}-e_{2}^{2})^2+l_{2}^{2}(l_{1}^{2}-e_{1}^{2})^2-2l_{1}l_{2}(l_{1}^{2}-e_{1}^{2})(l_{2}^{2}-e_{2}^{2})\cos\theta_{1}}}\right),
\nonumber\\
\chi_{2}&=&\cos^{-1}\left(\frac{e_{2}(l_{1}^{2}-e_{1}^{2})-e_{1}(l_{2}^{2}-e_{2}^{2})}{\sqrt{l_{1}^{2}(l_{2}^{2}-e_{2}^{2})^2+l_{2}^{2}(l_{1}^{2}-e_{1}^{2})^2-2l_{1}l_{2}(l_{1}^{2}-e_{1}^{2})(l_{2}^{2}-e_{2}^{2})\cos\theta_{1}}}\right). \label{eq:chi}
\end{eqnarray}
\end{widetext}
Therefore, if $\{q_i , \rho_i\}_{i=1}^{3}$ satisfies the conditions, $r_{1}^*$ becomes
$\frac{l_{1}^{2}-e_{1}^{2}}{2(l_{1}\cos\chi+e_{1})}$ and
the guessing probability $P_{\rm{guess}}$ is given by
\begin{equation}\label{eq:tqsgp}
P_{\rm{guess}}=q_{1} +
\frac{l_{1}^{2}-e_{1}^{2}}{2(l_{1}\cos\chi+e_{1})}.
\end{equation}
 The optimal POVM can be found by substituting $\{p_i ,\vec{w}_i  \}_{i=1}^3$
 into Eq.(\ref{eq:optm}). Through a lengthy calculation, we find $\{p_i ,\vec{w}_i \}_{i=1}^3$
 such as
\begin{eqnarray}
p_1 &=&\frac{l_1 l_2 \sin \theta_1 -r_1 l_1 \sin \chi -r_1 l_2 \sin (\theta_1 -\chi)}{l_1 l_2 \sin \theta_1 +e_2 l_1 \sin \chi + e_1 l_2 \sin (\theta_1 -\chi)},\nonumber\\
p_2 &=&\frac{r_2 l_2 \sin (\theta_1 -\chi)}{l_1 l_2 \sin \theta_1 +e_2 l_1 \sin \chi + e_1 l_2 \sin (\theta_1 -\chi)},\nonumber\\
p_3 &=&\frac{r_3 l_1 \sin \chi}{l_1 l_2 \sin \theta_1 +e_2 l_1 \sin \chi + e_1 l_2 \sin (\theta_1 -\chi)},
\end{eqnarray}
and,
\begin{eqnarray}
\vec{w}_1 &=& \frac{ \sin (\theta_1 -\chi)}{l_1\sin \theta_1 }(q_2 \vec{v}_2 -q_1 \vec{v}_1)+\frac{ \sin \chi}{l_2 \sin \theta_1 } (q_3 \vec{v}_3 -q_1 \vec{v}_1 ), \nonumber\\
\vec{w}_2 &=& \frac{r_1 \vec{w}_1 - (q_2 \vec{v}_2 -q_1 \vec{v}_1)}{r_1 + e_1},\nonumber\\
\vec{w}_3 &=& \frac{r_1 \vec{w}_1 - (q_3 \vec{v}_3 -q_1 \vec{v}_1)}{r_1 + e_2}.
\end{eqnarray}
However, the case where this condition is not satisfied turns out to be a problem of discriminating two quantum states. Therefore the guessing probability to the case can be given by corollary \ref{coro:ep2}.
 Now we can have lemma \ref{lem:tsd}.
\begin{lemma}[three quantum states discrimination]
When arbitrary three quantum states $\{q_{i},\rho_{i}\}_{i=1}^{3}$
are given with given priori probabilities, the guessing probability
can be classified into the following three cases: (i) When the number
of the extreme points to $P\{q_{i},\rho_{i}\}_{i=1}^{3}$ is one, the
guessing probability becomes $q_{1}$ by the corollary \ref{coro:ep1}. (ii) When
the number of the extreme points is two or three and the condition of
Eq.(\ref{eq:tqnsc}) cannot be satisfied, the guessing probability
can be found by the corollary \ref{coro:ep2}. (iii) When the number of the extreme
points is three and the condition of Eq.(\ref{eq:tqnsc}) is
satisfied, the guessing probability can be given by
Eq.(\ref{eq:tqsgp}). \label{lem:tsd}
\end{lemma}
 Here as an example let us consider the quantum discrimination of three symmetric
 quantum states. The symmetric property implies that for $\rho_{1}$, $\rho_{2}$, and $\rho_{3}$, $\tr\rho_{1}\rho_{2}=\tr\rho_{2}\rho_{3}=\tr\rho_{3}\rho_{1}$.
Their purity is assumed to be the same as
$\tr\rho_{1}^{2}=\tr\rho_{2}^{2}=\tr\rho_{3}^{2}\leq 1$. This symmetric condition can be expressed by
\begin{equation}
\vec{v}_{i} \cdot \vec{v}_{j}=\left\{
\begin{array}{ll}
r & (i=j),\\
\gamma & (i \neq j),
\end{array}
\right.
\end{equation}
where $\vec{v}_{i}$ is the Bloch vector of $\rho_{i}$.
 If their priori probabilities are the same
as $\frac{1}{3}$ ($q_{1}=q_{2}=q_{3}=\frac{1}{3})$, the guessing
probability $P_{\rm guess}$ becomes $\frac{1}{3} + r_{1}^*$. Since
$P\{q_{i}\vec{v}_{i}\}_{i=1}^{3}=P\{\vec{v}_{i}/3\}_{i=1}^{3}$ is
the equilateral triangle whose side is given by $\sqrt{2(t-s)}/3$
($t \equiv 1-\gamma$ and $s \equiv 1-r$), $\{r_{i}, \vec{w}_{i}\}_{i=1}^{3}$
satisfying the geometric KKT conditions naturally exists. The
circumradius of the triangle $P\{r_{i}\vec{w}_{i}\}_{i=1}^{3}$
becomes $\frac{1}{3}\sqrt{\frac{2(t-s)}{3}}$. Therefore, we find
\begin{equation}
r_{1}=r_{2}=r_{3}=\frac{1}{3}\sqrt{\frac{2(t-s)}{3}}.
\end{equation}
The guessing probability $P_{\rm guess}$ turns out to be
\begin{equation}
P_{\rm guess}=\frac{1}{3}\left(1+\sqrt{\frac{2(t-s)}{3}}\ \right),
\end{equation}
which agrees with the result in \cite{ref:sugimoto}.

 In conclusion, by considering the minimum-error quantum state discrimination as the complementarity problem, we obtained four geometric optimality conditions in the case of qubit geometry. We clearly showed that there is a relation between these  conditions and the optimal measurement. By these conditions and the intrinsic polytope for the discrimination problem, we can  provide a method to discriminate $N$
 qubit-mixed quantum states. We are also able to obtain the guessing probability and the optimal measurements. We applied these results to discriminating three-qubit mixed quantum states to show that discrimination for the three-qubit mixed quantum states can be classified by the geometric KKT conditions and the number of extreme points for the intrinsic polytope. The analytic expression of the guessing probability and the optimal measurement for three-qubit mixed quantum states was obtained. Furthermore, we have shown that for the special case of three symmetric quantum states, our result is consistent .

\section*{Acknowledgment}
We would like to thank Dr. Bae Joonwoo and an anonymous referee for
reading the paper and commenting. This work is supported by the
Basic Science Research Program through the National Research
Foundation of Korea, funded by the Ministry of Education, Science, and
Technology (KRF2011-0027142 and KRF2012-0008086).

\section*{Appendix}
 Suppose that two points $T$ and $T^{'}$, whose distance is $l$, are
 given in a two-dimensional plane. The points where the difference in
 distances between two points $T$ and $T^{'}$ becomes $e$ form a
 hyperbola. When the distance from these points to $T$($T^{'}$)
becomes $r$($r^{'}$), the hyperbola can be divided into two curves
$r^{'}-r=e$ and $r-r^{'}=e$. Let us denote the curve
$r^{'}-r=e$ as $C_{e}\{T,T^{'}\}$. The distance $r$ can be obtained
from the hyperbolic equation as follows:
\begin{figure}[!ht]
\centerline{\includegraphics[scale=0.49]{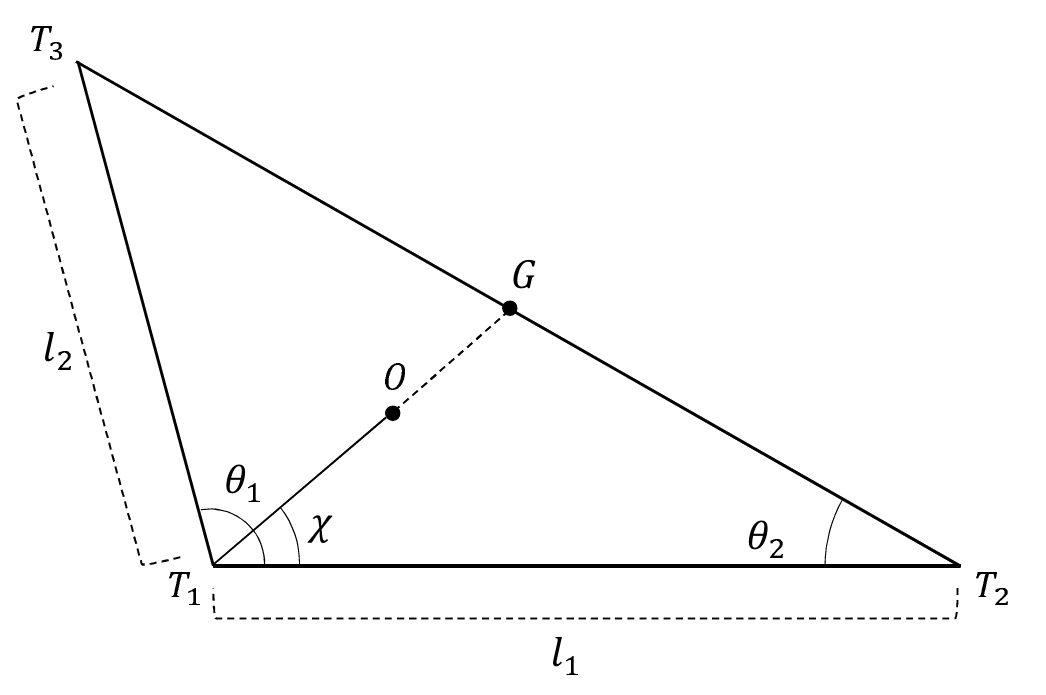}} \caption{ For the
point $O$ to be located inside the triangle, $\overline{T_{1}O}$ must be shorter than $\overline{T_{1}G}$.} \label{fig:tri}
\end{figure}
\begin{equation}\label{eq:hyper}
r=\frac{l^2 -e^2}{2(l \cos \theta +e)},
\end{equation}
 where $\theta$ is the angle between the segment to $r$ and the line segment $\overline{TT^{'}}$.
Now let us consider a triangle formed by three different points
$T_{1}$,$T_{2}$, and $T_{3}$ in a two-dimensional plane. We denote the
interior of the triangle, $C_{e_{1}}\{T_{1},T_{2}\}$, and
$C_{e_{2}}\{T_{1},T_{3}\}$ as $\triangle$, $C_{1}$, and $C_{2}$.
Also, let us represent the intersection of $\triangle$, $C_{1}$, and
$C_{2}$ as $\Omega$($\Omega=\triangle\cap C_{1} \cap C_{2}$). Now we
will find the necessary and sufficient condition where $\Omega$ is
nonempty. Since the condition for $\triangle\cap C_{i}$ to be
nonempty is $l_{i} > e_{i}$, we obtain the condition (i) of Eq. (24)
in the main text. Here $l_{1}$ and $l_{2}$ are the length of
$\overline{T_{1}T_{2}}$ and $\overline{T_{1}T_{3}}$, respectively.

 If the inner angle of the vertex $T_i$ is $\theta_{i}$, we can classify the
 triangle into four types, according to $\theta_{1}$:
 (i) $-\frac{e_{1}}{l_{1}},-\frac{e_{2}}{l_{2}} < \cos\theta_{1}$;
 (ii) $-\frac{e_{1}}{l_{1}} < \cos\theta_{1} \leq -\frac{e_{2}}{l_{2}} $;
 (iii) $-\frac{e_{2}}{l_{2}} < \cos\theta_{1} \leq -\frac{e_{1}}{l_{1}} $;
  (iv) $ \cos\theta_{1} \leq -\frac{e_{1}}{l_{1}}, -\frac{e_{2}}{l_{2}} $.
And the condition where $C_{1}\cap C_{2}$ becomes nonempty in each
case is as follows:
 (i) $\frac{l_{2}\cos\theta_{1} +e_{2}}{l_{2}+e_{2}} < \frac{l_{2} - e_{2}}{l_{1}-e_{1}} < \frac{l_{1}+e_{1}}{l_{1}\cos\theta_{1} +e_{1}}$,
 (ii) $\frac{l_{2} - e_{2}}{l_{1}-e_{1}} < \frac{l_{1}+e_{1}}{l_{1}\cos\theta_{1} +e_{1}}$,
 (iii) $\frac{l_{2}\cos\theta_{1} +e_{2}}{l_{2}+e_{2}} < \frac{l_{2} - e_{2}}{l_{1}-e_{1}}$, and
(iv) no condition needed. These conditions can be put into the two restrictive ones:
\begin{equation}
 \frac{l_1 \cos \theta_1 + e_1}{l_1 + e_1} < \frac{l_1 -e_1}{l_2 -e_2} ,\
 \frac{l_2 \cos \theta_1 + e_2}{l_2 + e_2} < \frac{l_2 -e_2}{l_1 -e_1},
\end{equation}which is the condition (ii) of Eq. (24) in the main text.
 Indeed, if $C_{1}$ and $C_{2}$ meet together, they intersect only at single point because the equation derived by Eq. (\ref{eq:hyper}),
\begin{equation}
\frac{l_{1}^{2}-e_{1}^{2}}{2(l_{1}\cos\chi+e_{1})}=\frac{l_2^2 -e_2^2}{2(l_2 \cos (\theta_1 -\chi)+e_2)},
\end{equation}
can be satisfied by unique $\chi\in (0,\theta_{1})$. When we denote
the intersection point as $O$, $\chi$ is $\angle OT_{1}T_{2}$, which
is given as follows:
\begin{widetext}
\begin{eqnarray}
\chi&=& \chi_{2} -\chi_{1}, \nonumber\\
\chi_{1}&=&
\cos^{-1}\left(\frac{l_{1}(l_{2}^{2}-e_{2}^{2})-l_{2}(l_{1}^{2}-e_{1}^{2})\cos\theta_{1}
}{\sqrt{l_{1}^{2}(l_{2}^{2}-e_{2}^{2})^2+l_{2}^{2}(l_{1}^{2}-e_{1}^{2})^2-2l_{1}l_{2}(l_{1}^{2}-e_{1}^{2})(l_{2}^{2}-e_{2}^{2})\cos\theta_{1}}}\right),
\nonumber\\
\chi_{2}&=&\cos^{-1}\left(\frac{e_{2}(l_{1}^{2}-e_{1}^{2})-e_{1}(l_{2}^{2}-e_{2}^{2})}{\sqrt{l_{1}^{2}(l_{2}^{2}-e_{2}^{2})^2+l_{2}^{2}(l_{1}^{2}-e_{1}^{2})^2-2l_{1}l_{2}(l_{1}^{2}-e_{1}^{2})(l_{2}^{2}-e_{2}^{2})\cos\theta_{1}}}\right). \label{eq:chi}
\end{eqnarray}
\end{widetext}
Here let us find the condition for $O\in
\triangle$.
 This can be found from the fact that
when $G$ is the intersection point between the half line from the
vertex $T_{1}$ to the point $O$ and the line segment
$\overline{T_{2}T_{3}}$, the length of $\overline{T_1 G}$ becomes
$\frac{l_{1} \sin \theta_{2}}{\sin(\chi+\theta_{2})}$. From Fig. \ref{fig:tri} we can see that the point $O$ can be located
inside the triangle if the length of $\overline{T_1 O}$ becomes less
than that of $\overline{T_1 G}$:
\begin{equation}
\frac{l_1^2 -e_1^2}{2(l_1 \cos \chi + e_1)} < \frac{l_1 \sin \theta_2}{\sin (\chi + \theta_2)}.
\end{equation}
Therefore we showed that three conditions given by Eq. (24) in the
main text are the necessary and sufficient conditions for nonempty
$\Omega$.

\end{document}